\documentclass[aps,twocolumn,superscriptaddress,10pt,floatfix]{revtex4-2}

\usepackage{graphicx}
\usepackage{dcolumn}
\usepackage{bm}
\usepackage{hyperref}
\hypersetup{
    unicode=false,          
    pdftoolbar=false,        
    pdfmenubar=true,        
    pdffitwindow=false,     
    pdfstartview={FitH},    
    pdfkeywords={Quantum batteries} {Ergotropy} {Micromasers}{Steady State}, 
    pdfnewwindow=true,      
    colorlinks=true,       
    linkcolor=black,          
    citecolor=blue,        
    filecolor=magenta,      
    urlcolor=blue           
}

\usepackage{xcolor}
\usepackage{mathrsfs}
\usepackage{amsfonts}
\usepackage{amsmath}
\usepackage{amssymb}
\usepackage{subfigure}
\usepackage{soul}

\newcommand{\e}{{\rm e}}
\newcommand{\imai}{{\rm i}}

\usepackage{color}

\newcommand{\be}{\begin{equation}}
\newcommand{\bw}{\begin{widetext}}
\newcommand{\ew}{\end{widetext}}
\newcommand{\bea}{\begin{eqnarray}}
\newcommand{\bc}{\begin{center}}            
\newcommand{\ee}{\end{equation}}
\newcommand{\eea}{\end{eqnarray}}
\newcommand{\ec}{\end{center}}

\begin{document}


\title{Thermodynamic uncertainty relation in nondegenerate and degenerate maser heat engines }

\author{Varinder Singh}
\email{varinder@ibs.re.kr}
\affiliation{Center for Theoretical Physics of Complex Systems, Institute for Basic Science (IBS), Daejeon - 34126, Korea}

\author{Vahid Shaghaghi}
\affiliation{Center for Theoretical Physics of Complex Systems, Institute for Basic Science (IBS), Daejeon - 34126, Korea}
\affiliation{Center for Nonlinear and Complex Systems, Dipartimento di Scienza e Alta
Tecnologia, Universit\`{a} degli Studi dell’Insubria, via Valleggio 11, 22100 Como, Italy}
\affiliation{
Istituto Nazionale di Fisica Nucleare, Sezione di Milano, via Celoria 16, 20133 Milano, Italy}

\author{\"{O}zg\"{u}r E. M\"{u}stecapl{\i}o\u{g}lu}
\affiliation{Department of Physics, Ko\c{c} University, 34450 Sar\i{}yer, Istanbul, Turkey}
\affiliation{TÜBİTAK Research Institute for Fundamental Sciences, 41470 Gebze, Turkey}

\author{Dario Rosa}
\affiliation{Center for Theoretical Physics of Complex Systems, Institute for Basic Science (IBS), Daejeon - 34126, Korea}
\affiliation{Basic Science Program, Korea University of Science and Technology (UST), Daejeon - 34113, Korea}

\date{\today}

\begin{abstract}

We investigate the thermodynamic uncertainty relation (TUR), \textit{i.e.} a trade-off between entropy production rate and relative power fluctuations, for nondegenerate three-level and degenerate four-level maser heat engines. In the nondegenerate case, we consider two slightly different configurations of three-level maser heat engine and contrast their degree of violation of standard TUR. In the high-temperature limit, standard TUR relation is always violated for both configurations. We also uncover an invariance of TUR under simultaneous rescaling of the matter-field coupling constant and system-bath coupling constants.  For the degenerate four level engine, we study the effects of noise-induced coherence on TUR. We show that depending on the parametric regime of operation, noise-induced coherence can either suppress or enhance the relative power fluctuations.

\end{abstract}

\maketitle
\section{Introduction}
Since the dawn of industrial revolution, heat engines have played an important role in the development of classical thermodynamics on both  theoretical and  experimental fronts. Even with the rise of quantum thermodynamics \cite{Kosloff2013,Sai2016,Sourav2021,Mahler,AlickiKosloff,DeffnerBook,Asli2020} --- the study of heat and work in the quantum regime --- heat engines remain among the central research topics. In order to harness quantum resources such as entanglement, coherence and quantum fuels to our advantage (which is at the heart of the flourishing area of quantum technologies), it is important to understand the energy conversion process at nanoscale. However, thermal machines operating at nanoscale are subject to thermal as well as quantum fluctuations due to their small size, which may negatively affect their performance. Hence, it is crucial to develop a fundamental understanding of these fluctuations characterizing the performance of quantum thermal machines.   
Recently, a conceptual advance has been made in this direction by Barato and coauthors \cite{SeifertTUR} by introducing the thermodynamic uncertainty relation (TUR). In the context of  steady state classical heat engines, the TUR states that there is always trade-off between the relative fluctuation in the power output and thermodynamic cost (rate of entropy production $\sigma$) of maintaining  the nonequilibrium steady state:
\begin{equation}
\mathcal{Q}\equiv \sigma \frac{\text{Var}(P)}{P^ 2}\ge 2, \label{TUR}
\end{equation}
where  $P$ and $\text{Var}(P)$ represent the mean and variance of the power in the steady state. 
Eq. (\ref{TUR}) was originally discovered for biomolecular processes \cite{SeifertTUR} and proved by using the formalism of large deviation theory \cite{Gingrich2016}.

The original TUR, Eq.(\ref{TUR}), which we will refer to as standard thermodynamic uncertainty relation (STUR) from now on, is applicable to  systems in nonequilibrium steady-state obeying a Markovian continuous-time dynamics with explicit  time-independent driving \cite{Horowitz2020}. Later, it was shown to hold in finite-time \cite{Pietzonka2017,Horowitz2017}. Without any one of the assumptions mentioned above, Eq. (1) can be violated \cite{Horowitz2020}. Thus, a number of generalizations have been proposed in various settings \cite{PietzonkaBarato2016,Polettini2016,Proesmans2017,Shiraishi2017,PietzonkaSeifert2016,Pigolotti2018,Barato_2018,Pigolotti2017,Fischer2018,Dechant2018,Dechant_2018,Barato2016prx,Macieszczak2018,Koyuk2020,LeePark2019,Garrahan2017,VuHasegawa2019,LiuUeda2020,Gingrich2017,Terlizzi,LeePark2021,JongMin2021,Busiello2022, francica2022fluctuation, carrega2019optimal, cangemi2021optimal, cangemi2020violation}. 
Additionally, there have been considerable amount of efforts to probe the validity and extensions of STUR in quantum systems  \cite{Krzysztof,Bijay2018,Saryal2021,LiuSegal2019,Guarnieri2019,Patrick2019,Tilmann2021,Gerry2022,LiuSegal2021,Mark2021,Timpanaro2021A,Timpanaro2021B,Davinder2021,Menczel2021,Kazutaka2021,BenentiWang2020,Kerremans2022,Miller2021,Souza2022,Saryal2021B,Hasegawa2020,Hasegawa2019,Timpanaro2019,Carollo2019}

Recently, the role of quantum coherence  in the violations of TURs have been explored in details \cite{Patrick2021,VuSaito2022}. Specifically, it has been shown that fluctuations are not encoded in the steady  state alone and STUR violations can be seen as the consequence of coherent dynamics going beyond steady-state coherence  \cite{Patrick2021,VuSaito2022}.  
In these papers, the attention is put on the effect of drive-induced coherence on TURs. On the other hand, the effect of \textit{noise-induced coherence} on TURs is usually left apart, with just a couple of notable exceptions (which actually refer to models of quantum absorption refrigerators) \cite{LiuSegal2021,Holubec2019}.
We recall that the phenomenon of noise-induced coherence -- arising due to interference between different transition paths  from the degenerate energy levels to a common level -- has shown to drastically increase the power output of quantum heat engines \cite{Scully2011,Dorfman2018}. Therefore, it is interesting to explore the effects of noise-induced coherence on the  power fluctuations in a quantum heat engine.


In this work, we explore the role of quantum coherence, both  drive-induced and noise-induced, in the violations of STUR in different variants of maser heat engines, introduced by Scovil and  Schulz-Dubois (SSD) back in 1959 \cite{Scovil1959}. 
The SSD engine converts the incoherent thermal energy of  heat reservoirs into a coherent maser output \cite{Scovil1959,VJ2019,Varinder2020,Geva1994,Geva1996,BoukobzaTannor2007,BoukobzaTannor2006A,VJ2020,Kiran2021} and it is one of the very few experimentally realizable quantum heat engines \cite{Klatzow2019}.

Given its prominence both experimentally and theoretically, it is therefore of great interest to curb fluctuations in the  power output of the SSD engine, thus motivating a detailed study of the model in the context TURs.

The paper is organized as follows. In section \ref{sec:model_def} we review the SSD model. In Section \ref{sec:TUR_3_level} we study the violations of STUR for two different variants of the SSD model and compare their respective degrees of violation.
In Section \ref{sec:SSD_4_level_model} we introduce a variant of the SSD model having two degenerate levels, leading to a scenario suitable to study the effect of noise-induced coherence on violations of STUR. Such an analysis is then performed in Section \ref{sec:TUR_4_level}. In section \ref{sec:conclusions} we conclude our paper.
  \begin{figure*}   [t]
 \begin{center}
 \includegraphics[trim=0cm 0cm 0cm 0cm, angle=0, width=1\textwidth]{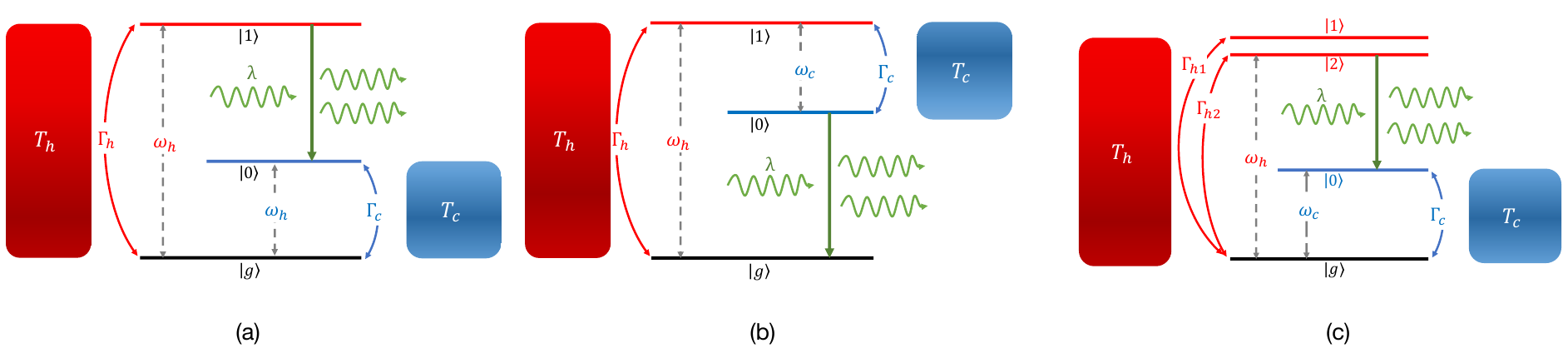}
 \end{center}
\caption{(a) Model I of three-level laser heat engine continuously coupled to two reservoirs of temperatures
$T_h$ and $T_c$ having coupling constants $\Gamma_h$ and $\Gamma_c$, respectively. The system is interacting with a
classical single mode field. $\lambda$ represents the strength of matter-field coupling. (b) Model II is slightly different from the Model I. Here, cold reservoir
is connected to two upper levels instead of two lower levels as in Model I. Similarly, power mechanism is coupled to lower two levels instead of upper two levels. (c) Four-level degenerate maser heat engine having degeneracy in the upper level.}
\end{figure*} 

\section{The SSD model}
\label{sec:model_def}
The SSD engine \cite{Scovil1959} is one of the most well-known examples of quantum heat engines. In this model, a three-level system is simultaneously 
coupled to two thermal reservoirs at different temperatures $T_c$ and $T_h$ ($T_c<T_h$) (see Fig. 1(a)). The hot reservoir supplies heat to  induce transition between the  states $\vert g\rangle$ and  $\vert 1\rangle$, whereas the cold reservoir de-excites
the transition between the  states  $\vert 0\rangle$ and $\vert g\rangle$. The power output mechanism between states $\vert 0\rangle$ and
$\vert 1\rangle$ is modeled by coupling the transition between them  to a single-mode classical  field.  
$H_0=\hbar \sum \omega_j \vert j\rangle\langle k\vert$ is the free Hamiltonian of the system, where $\omega_j$'s represent the atomic frequencies.
The following semiclassical Hamiltonian  describes the interaction between 
the system and the classical field of frequency $\omega$ in the rotating wave approximation:
$V(t)=\hbar \lambda (e^{-i\omega t} \vert 1\rangle\langle 0\vert + e^{i\omega t} \vert 0\rangle\langle 1\vert)$; 
$\lambda$ is the field-matter coupling constant. In a reference frame rotating with respect to system Hamiltonian $H_0$, the 
dynamics of the three-level system is described by the following   Lindblad master equation:
\begin{equation}
\dot{\rho} = -\frac{i}{\hbar} [V_R,\rho] + \mathcal{L}_{h}[\rho] + \mathcal{L}_{c}[\rho],
\end{equation}
where, $\rho$ is the density matrix for the three-level system, the coherent part of the dynamics is controlled by $V_R=\hbar\lambda(\vert 1\rangle\langle 0\vert + \vert 0\rangle\langle 1\vert)$ and $\mathcal{L}_{h(c)}[\rho]$ describes  the interaction between the system and the hot (cold) reservoir. 
In details, we have
\begin{eqnarray}
\mathcal{L}_h[\rho] &=& \Gamma_h(n_h+1) \big (\sigma_{g1} \rho \sigma^{\dagger}_{g1} -\frac{1}{2}  \{\sigma^{\dagger}_{g1} \sigma_{g1},\rho\}   \big) \nonumber
\\
&& +\Gamma_h n_h  \big (\sigma^{\dagger}_{g1} \rho \sigma_{g1} -\frac{1}{2}  \{\sigma_{g1} \sigma^{\dagger}_{g1} ,\rho\}   \big) , \label{D1}
\end{eqnarray} 
\begin{eqnarray}
\mathcal{L}_c[\rho] &=& \Gamma_c(n_c+1) \big (\sigma_{g0} \rho \sigma^{\dagger}_{g0} -\frac{1}{2}  \{\sigma^{\dagger}_{g0} \sigma_{g0},\rho\}   \big) \nonumber
\\
&& +\Gamma_c n_c  \big (\sigma^{\dagger}_{g0} \rho \sigma_{g0} -\frac{1}{2}  \{\sigma_{g0} \sigma^{\dagger}_{g0} ,\rho\}   \big) , \label{D2}
\end{eqnarray}
where $\sigma_{gk}=\vert g\rangle \langle k\vert$, $k=0, 1$. $\Gamma_c$ and $\Gamma_h$ are system-bath coupling constants for cold and hot reservoirs, respectively.  
$n_{h(c)}= 1/(\exp[\hbar\omega_{h}/k_B T_{h}]-1)$ and  $n_{c}= 1/(\exp[\hbar\omega_{c}/k_B T_{c}]-1)$ represent average number of photons with mode frequencies 
$\omega_h$ and $\omega_c$ in the hot and cold reservoirs, respectively (
 $\omega_c=\omega_0-\omega_g$, $\omega_h=\omega_1-\omega_g$).

%

\section{TUR for the SSD model}
\label{sec:TUR_3_level}
In this section we analyze the SSD model from the viewpoint of TUR and we will compare two slightly different implementations of it.
These two implementations differ by which energy levels in the three-level systems are connected by the cold reservoir: in the first implementation the cold reservoir connects $|g\rangle$ with $| 0 \rangle$ while in the second the cold reservoir connects $|0\rangle$ with $| 1 \rangle$.
These two configurations are often considered interchangeable in the literature and they are both referred to as the SSD model, \cite{Scovil1959,Geva1994,Geva1996,BoukobzaTannor2007,BoukobzaTannor2006A,VJ2019,Klatzow2019}.
However, as we will discuss in the following, from the point of view of TUR these two configurations are rather different. 
TUR in the first implementation has not been analyzed before while the second implementation is the focus of Refs.\cite{Patrick2021,VuSaito2022}.

\subsection{Model I}
First, we will investigate the TUR in Model I as shown in Fig. 1(a). The TUR quantifier $\mathcal{Q}=\sigma \text{Var}(P)/P^2$ is evaluated using the method of full counting statistics; see Appendix C for details. The calculations yield the following result 
\begin{widetext}
\begin{equation}
\mathcal{Q}^{\rm I}(\Gamma_h, \Gamma_c, \lambda, n_h, n_c) = \frac{1}{A(n_h - n_c)}
\Bigg[          
		A(n_h + n_c + 2 n_h n_c) +  \frac{8(n_h-n_c)^2 \, \lambda^2\,\Gamma_c \Gamma_h }{A \,B \, \Gamma_c \Gamma_h + C\,\lambda^2}
		\Bigg(2-\frac{D + F + G + H}{A \,B \, \Gamma_c \Gamma_h + C\,\lambda^2}\Bigg)
\Bigg]  \ln\left[ \frac{n_h(n_c+1)}{n_c(n_h+1)}  \right],     \label{TURme} 
\end{equation}
where 
$A = \Gamma_c(1+n_c) + \Gamma_h(1+n_h)$, $B = 1+2n_h+n_c(2+3n_h)$, 
$C = 4\,[\Gamma_c(1+3n_c) + \Gamma_h(1+3n_h)]$,  
$D =(1+2n_c)\Gamma_c \big[ (1+n_c)^2 \, \Gamma_c^2 + 16\lambda^2 \big]$,
$F = (1+2n_h)\Gamma_h \big[ (1+n_h)^2 \, \Gamma_h^2 + 16\lambda^2 \big]$, 
$G = (1+n_c)[7 + 13n_c + 6 (2+3n_c)] \Gamma_c^2 \Gamma_h$, 
$H = (1+n_h)[7 + 13n_h + 6 (2+3n_h)] \Gamma_h^2 \Gamma_c$.        We note that the first term, $\mathcal{Q}_{\rm pop}= \ln \{n_h(n_c+1)/n_c(n_h+1)\}(n_h+n_c+2n_h\,n_c)/(n_h-n_c)$, given in Eq. (\ref{TURme})   depends on the bath populations only.       \end{widetext} 
 Using the inequalities $a/(a-b)\ln(a/b)\ge 1$ and $b/(a-b)\ln(a/b)\ge 1$, we can show that $\mathcal{Q}_{\rm pop}\ge 2$. Thus we can associate the possible violations of STUR to negative values of remaining terms in Eq. (\ref{TURme}). 
 We also notice that at the verge of the population invertion (threshold condition for the masing),  $n_h=n_c$. Thus,  we have $\mathcal{Q}=2$, \textit{i.e.} STUR is saturated. 
 Finally, we numerically studied Eq. (\ref{TURme}) outside the equilibrium condition and for various values of the parameters. As an example, in  Fig. 2 (dashed blue curve) we report Eq. (\ref{TURme}) as a function of matter-field coupling strength $\lambda$ for fixed values of the other parameters. We note very weak violation of standard TUR for certain range of parameter $\lambda$.  We checked that upon further increasing of $\lambda$, $\mathcal{Q}$ gets saturated (not shown). Here, we want to emphasize that our engine always works in weak coupling regime ($\lambda/\omega\ll 1$).
 
 To complement the discussion on TUR, we would like to talk about another important quantity which becomes relevant when we take into account the fluctuations in the heat engines, the so-called reliability $\mathcal{R}$. Reliability of an engine is defined as the ratio between the mean power output and the square root of the variance of the power \cite{DingTalkner2018,Jeongrak2021,Shastri2022,Alam2022},
\be
\mathcal{R} = \frac{P}{\sqrt{\text{Var}(P)}}.
\ee
We have plotted reliability of the Model I in the inset of Fig. 2 represented by dashed blue curve, which shows that the initially reliability $\mathcal{R}^{\rm I}$ increases with increasing $\lambda$ and then becomes saturated.

Further, by inspecting Eq. (\ref{TURme}), we can highlight certain other interesting features of the TUR for the model under consideration.  We note that the TUR quantifier $\mathcal{Q}$ remains invariant if we scale system-bath and matter-field coupling constants by the same factor. In other words, simultaneous transformations of $\Gamma_c\rightarrow k \Gamma_c$, $\Gamma_h\rightarrow k \Gamma_h$ and $\lambda\rightarrow k \lambda$ leave $\mathcal{Q}$ invariant. Mathematically,
\be
\mathcal{Q}^{\rm I}(k \Gamma_h, k \Gamma_c, k \lambda, n_h, n_c)  =\mathcal{Q}^{\rm I}(\Gamma_h, \Gamma_c, \lambda, n_h, n_c).
\ee
On the other hand, such a scaling property does not hold for the reliability $\mathcal{R}$ as well as for the single quantities entering in STUR, i.e. for $P$, $\sigma$ and $\mathrm{Var}(P)$ individually.

Now, we turn our attention to the  ultra high-temperature limit, which is considered to be the classical limit.  In the ultra high-temperature limit, Eq. (\ref{TURme})  can be further simplified. In this regime, we can approximate $n_{h(c)}=k_B T_{h(c)}/\hbar \omega_{h(c)}\gg 1$. Then, Eq. (\ref{TURme})  reduces to the following form,
\begin{widetext} 
\be
\mathcal{Q}_{\rm HT} =  
  2- \frac{16(n_h-n_c)^2\Gamma_h\Gamma_c\lambda^2 \big(\Gamma_c^2 n_c^2+\Gamma_h^2 n_h^2+ 5\Gamma_c\Gamma_h n_h n_c+\lambda^2\big)}{9 n_h n_c (\Gamma_c n_c+\Gamma_h n_h)^2(4\lambda^2+\Gamma_h\Gamma_c n_h n_c)^2}. \label{TURHT}
\ee
\end{widetext} 
It is clear from Eq. (\ref{TURHT}) that unless $n_h=n_c$, $\mathcal{Q}_{\rm HT}$ is always smaller than 2, which implies that STUR is always violated in the maser heat engine operating in the ultra high temperature regime.  The STUR violations can be thought of arising from the coherent quantum dynamics which goes beyond the steady state coherences as shown in the Refs. \cite{Patrick2021,VuSaito2022}.
\begin{figure}   
 \begin{center}
\includegraphics[width=8.6cm]{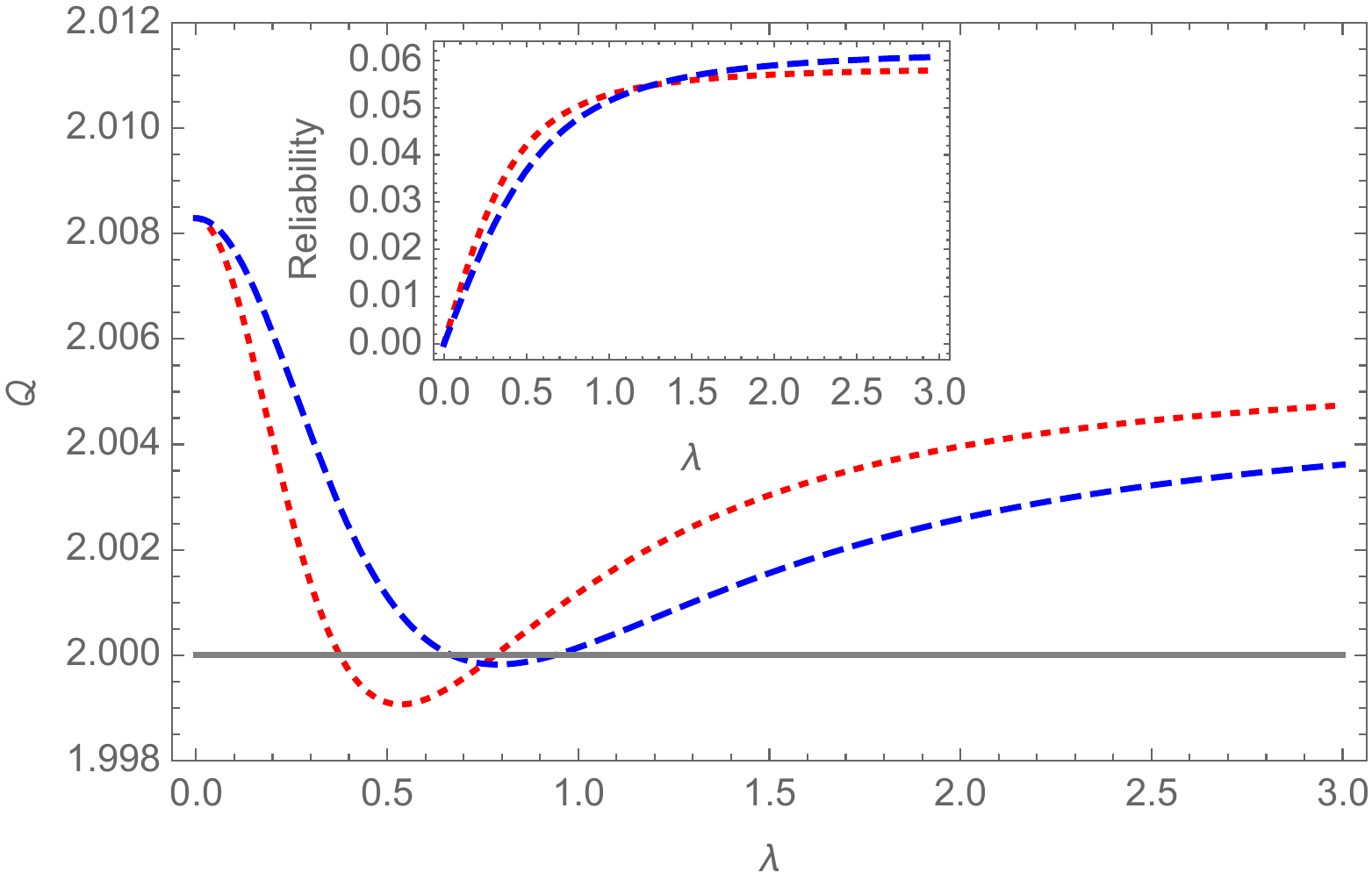}
 \end{center}
\caption{TUR quantifier $\mathcal{Q}$ versus matter-field coupling parameter $\lambda$. Dashed blue and dotted red curves correspond to Model I and Model II, respectively. Here, $\Gamma_h=0.1$, $\Gamma_c=2$, $n_h=5$, $n_c=0.027$.  In the inset, we have plotted reliability $\mathcal{R}$ of the engine versus matter-field coupling parameter $\lambda$. Dashed blue and dotted red curves correspond to Model I and Model II, respectively.  
}
\end{figure}
\subsection{Model II}
In this subsection, we consider a slight modification of Model I, which we refer as Model II and which is depicted in Fig. 1 (b). Certain aspects of TUR violations in this model have already been discussed in Ref. \cite{Patrick2021}. Here and in the next subsection, we will show that, although the two configurations are very similar (and they are often both referred as the SSD model) they give quite different results when dealing with violations of TUR. We will also show that this difference is coming from the spontaneous emission contribution which is not symmetric in the two configurations. 

For the Model II, we have the following expression for the TUR quantifier $\mathcal{Q}^{\rm II}$ ,
%
\bw
\be
\mathcal{Q}^{\rm II} =  \ln\left[ \frac{n_h(n_c+1)}{n_c(n_h+1)}  \right]  \Bigg\{ \frac{n_h+n_c+2n_h n_c}{n_h-n_c} + \frac{8(n_h-n_c)\Gamma_c\,\Gamma_h\,\lambda^2}{A'(A'\,B'+4 D'\,\lambda^2)}
\left[     2 - \frac{A(4B + A\, C)+16C\lambda^2}       {A'(A'\,B'+4 D'\,\lambda^2)}      \right] \Bigg\}, \label{TURpatrick}
\ee
where $A' = \Gamma_c n_c  +  \Gamma_h n_h$,  $B' =   n_c  +   n_h + 3n_c\,n_h$,  $C' = \Gamma_c (1+2n_c)  +  \Gamma_h (1+2n_h)$, 
$D' = \Gamma_c (2+3n_c)  +  \Gamma_h (2+3n_h)$.  \ew  The TUR relation for Model II given in Eq. (\ref{TURpatrick}) is different from the TUR relation given in Eq. (\ref{TURme}) for Model I. We plot Eq. (\ref{TURpatrick}) (dotted red curve) as well as reliability ($\mathcal{R}^{II}$) (dotted red curve in the inset of Fig. 2) of the engine as functions of matter-field coupling constant $\lambda$. Also for this model, we see  violation of STUR for certain values of $\lambda$. 

\subsection{Comparison between Model I and Model II}

Before we proceed further, a few comments about the similarities and differences between Model I and Model II are in order. 
As clear from Fig. 2 (a), Model II yields lower TUR ratio for smaller values of $\lambda$ as compared to the Model I (solid blue curve) whereas for the larger values of  $\lambda$, Model II yields better results. 
%
%
The converse analysis is true for the reliability of the engine (see inset of Fig. 2).   For the relatively smaller values of $\lambda$, Model II outperforms Model I whereas for relatively higher values of $\lambda$, Model I yields better reliability than Model II.

Although both models yield different expressions for the TUR ratio, the first term,  $\mathcal{Q}_{\rm pop}= \ln \{n_h(n_c+1)/n_c(n_h+1)\}(n_h+n_c+2n_h\,n_c)/(n_h-n_c)$,  is same. $\mathcal{Q}_{\rm pop}$ depends only on the bath populations, and is always greater than 2 \cite{Patrick2021}. Hence, in both models, the second term is responsible for the observed STUR violations. Although STUR violations are not uncommon in our three-level models, still they adhere to the quantum mechanical bound found in Ref. \cite{Guarnieri2019}, which is two times looser than the STUR. Further, in the high-temperature regime, Eq. (\ref{TURpatrick}) reduces to Eq. (\ref{TURHT}). Hence, in the high-temperature limit, both models lead to the same TUR ratio. This can be traced back to the identical dynamical rate equations for both models. 

In the high-temperature limit spontaneous emission can be ignored. Therefore the asymmetry between the two models, due to the presence of spontaneous emission, is no longer there.
In such a case, Model I and Model II share a reflection symmetry, thereby yielding the identical results. Mathematically, this can be seen as follows. Interchanging  the indices $g\rightarrow 1$ in Eqs. (A6)-(A10) and ignoring $\Gamma_h$ as compared to $\Gamma_h n_h$, we obtain the exactly same set of equations as given in Eqs. (A1)-(A5).

To make the comparison between two models more concrete, we also plot the histograms of sampled values of $\mathcal{Q}^{\rm I}$ and $\mathcal{Q}^{\rm II}$ for randomly sampling over a region of the parametric space (see Fig. 3). In both cases, for the great majority of sampled operational points, the TUR ratio stays close to the convention STUR limit $\mathcal{Q}=2$. However, it is clear from the histograms in Fig. 3 that STUR violations are more common in Model II. Additionally, as far as the minimum numerical value of TUR ratio is concerned, Model II attains lower minimum value as compared to Model I, \textit{i.e.} $\mathcal{Q}^{\rm II}_{\rm min}<\mathcal{Q}^{\rm I}_{\rm min}$.

Summarizing the results of this section, we presented a clear case where the physics of TUR is highly controlled by the spontaneous emission phenomenon: the two models just differ by the spontaneous emission term and this small difference makes them inequivalent in the way in which they violate the STUR. 

	\begin{figure*}
		\centering
		\includegraphics[width=1\textwidth]{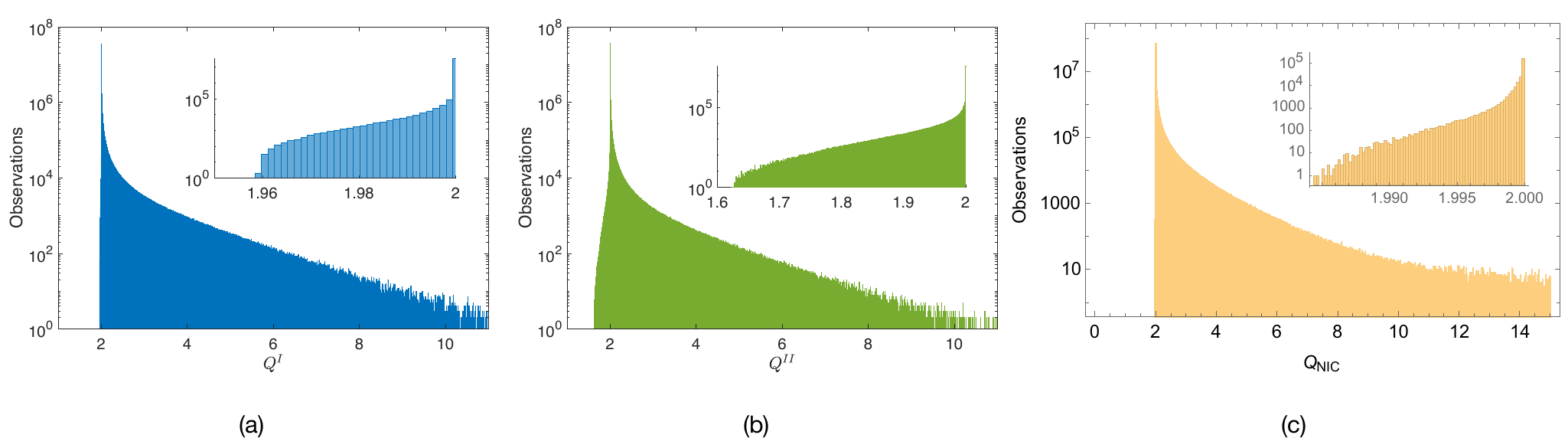}
		\caption{Histograms of sampled values of $\mathcal{Q}^{\rm I}$ and $\mathcal{Q}^{\rm II}$ for randomly sampling over a region of the parametric space.  The insets show the subset of the sampled data for which STUR violations are happening. The parameters are sampled over the uniform distributions $\Gamma_{h, c}\in [10^{-4},5]$, $n_{h, c}\in [0,10]$ and $\lambda\in[10^{-4},1]$. For plotting the histograms, we choose bin width of 0.01 to arrange $10^8$ data points.}
	\end{figure*}	
\section{A four-level version of the SSD model}
\label{sec:SSD_4_level_model}

In this section we consider a variant of the SSD model (see Fig. 1 (c)), originally introduced in \cite{Scully2011}. In this model, the upper levels $\vert 1\rangle$ and $\vert 2\rangle$  are degenerate. The Hamiltonian of the system in the rotating wave approximation is given by: 
$H_0=\hbar \sum \omega_k \vert k\rangle\langle k\vert$ where the summation runs over all four states. The interaction Hamiltonian takes the following form: 
$V(t)=\hbar \lambda e^{-i\omega t} \left(\vert 1\rangle\langle 0\vert 
+
\vert 2\rangle\langle 0\vert \right) + h.c.$. The time  evolution of the system is described by the following  master equation:
\begin{equation}
\dot{\rho} = -\frac{i}{\hbar} [H_0+V(t),\rho] + \mathcal{L}_{h}[\rho] + \mathcal{L}_{c}[\rho],
\end{equation}
where $\mathcal{L}_{h(c)}$ represents the dissipative Lindblad superoperator describing the system-bath 
interaction with the hot (cold) reservoir:
\begin{widetext}
\begin{eqnarray}
\mathcal{L}_c[\rho]= \Gamma_c(n_c+1) \Big(A_c\rho A_c^\dagger  -\frac{1}{2}\big\{ A_c^\dagger A_k,\rho \big\}\Big) 
+  \Gamma_c n_c\Big(A_c^\dagger \rho A_c  -\frac{1}{2}\big\{ A_c A_c^\dagger,\rho \big\}\Big)  , \label{disscold}
\end{eqnarray}
\begin{eqnarray}
\mathcal{L}_h[\rho]=\sum_{k={1,2}}\Gamma_{hk}\Big[(n_h+1)\Big(A_k\rho A_k^\dagger  -\frac{1}{2}\big\{ A_k^\dagger A_k,\rho \big\}\Big) 
+  n_h\Big(A_k^\dagger \rho A_k  -\frac{1}{2}\big\{ A_k A_k^\dagger,\rho \big\}\Big)\Big] \nonumber
\\
+  \Gamma\cos\theta \Big[(n_h+1)\Big(A_1\rho A_2^\dagger -\frac{1}{2}\big\{ A_2^\dagger A_1,\rho \big\}\Big) 
 +
n_h\Big(A_1^\dagger \rho A_2  -\frac{1}{2}\big\{ A_2 A_1^\dagger,\rho \big\}\Big)\Big] \nonumber
\\
+  \Gamma\cos\theta \Big[(n_h+1)\Big(A_2\rho A_1^\dagger -\frac{1}{2}\big\{ A_1^\dagger A_2,\rho \big\}\Big) 
 +
n_h\Big(A_2^\dagger \rho A_1  -\frac{1}{2}\big\{ A_1 A_2^\dagger,\rho \big\}\Big)\Big] \label{disshot}
\end{eqnarray} 
\end{widetext}
where  $A_c=\vert g\rangle \langle 0\vert$,  
$A_k=\vert g\rangle \langle k \vert$ ($k=1, 2$) are jump operators between the relevant transitions. $\theta$ is the angle between the dipole transitions $\vert 1\rangle\rightarrow \vert g\rangle$    and $\vert 2\rangle\rightarrow \vert g\rangle$. $\Gamma=\sqrt{\Gamma_{h1}\Gamma_{h2}}$, where $\Gamma_{h1}$ and $\Gamma_{h2}$ are Wigner-Weisskopf constants for transitions between  $\vert g\rangle\rightarrow \vert 1\rangle$ and $\vert g\rangle\rightarrow \vert 2\rangle$, respectively. 
To make the discussion analytically traceable, from now on we set  $\Gamma_{h1}=\Gamma_{h2}=\Gamma_h$, thus we have $\Gamma=\Gamma_h$.

Physically, the phenomenon of noise-induced coherence arises due to the interference of two indistinguishable decay paths $\vert 1\rangle\rightarrow \vert g\rangle$ and $\vert 2\rangle\rightarrow \vert g\rangle$ to the same level $\vert g\rangle$ \cite{Scully2011}. As customary when dealing with noise-induced coherence, we define $p\equiv\cos\theta$ as the noise-induced coherence parameter, lying in the range $(-1, 1)$. For the values of $p$ lying between (p, -1, 0), destructive interference takes place between the dipole transitions whereas in the range $(p, 0, 1)$, dipole transitions interfere constructively.

\section{TUR in four-level degenerate maser heat engine}
\label{sec:TUR_4_level}
\begin{figure}   
 \begin{center}
\includegraphics[width=8.6cm]{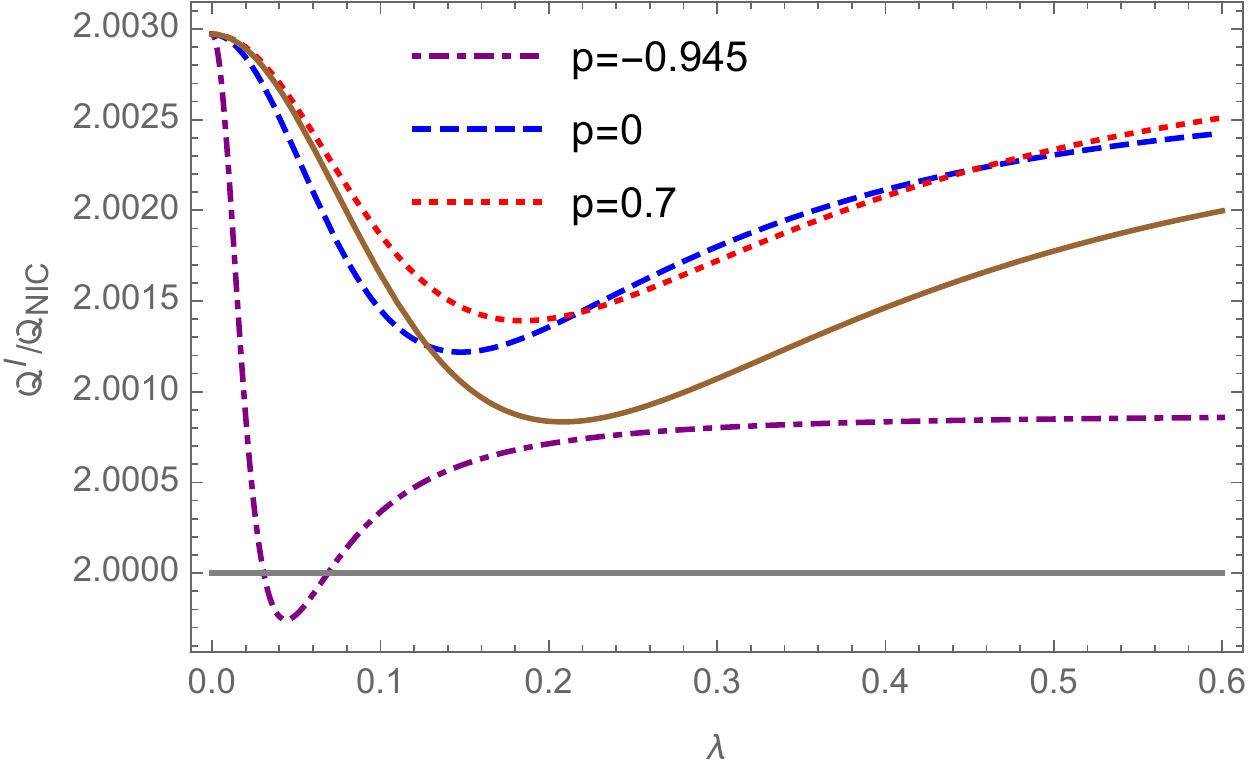}
 \end{center}
\caption{TUR quantifier  $\mathcal{Q}_{\rm NIC}$    as a function of $\lambda$ for different values of noise-induced coherence parameter $p$. Dot-dashed violet curve, dashed blue curve and dotted red curve represent the cases for $p=-.945$, $p=0$ and $p=0.7$, respectively. Solid brown curve represents the TUR ratio for the three-level engine, Model I. Here, $\Gamma_h=0.3$, $\Gamma_c=0.03$, $n_h=6$ and $n_c=3$.
}
\label{FigQNIC}
\end{figure}

In this case, the resulting form of TUR quantifier $\mathcal{Q}_{\rm NIC}$ is very complicated and not at all illuminating. Therefore, we will present the numerical results only. $\mathcal{Q}_{\rm NIC}$  is plotted in Fig. \ref{FigQNIC} as a function of $\lambda$ for different values of noise-induced coherence parameter $p$.  Solid purple curve in Fig. {\ref{FigQNIC}} represents the TUR ratio for the three-level engine, Model I. It is clear from  Fig. {\ref{FigQNIC}} that   depending upon the numeric value of noise-induced coherence parameter $p$, 
 noise-induced coherence can either suppress or enhance the relative fluctuations in the power output of degenerate four-level engine as compared to its three-level counterpart. For $p=-0.945$, the dot-dashed violet curve always lies below the solid brown curve for three-level engine, thereby showing the advantage of noise-induce coherence in suppressing the relative power fluctuations. For $p=0$,  dashed blue curve representing  $Q_{\rm NIC}(p=0)$ lies below   its three-level counterpart (solid brown curve) for smaller values of $\lambda$. However, for relatively higher values of $\lambda$, noise-induced coherence is not helpful and enhances the relative power fluctuations in the engine. For $p=0.7$, we have a different story altogether. In this case, dotted red curve (representing the case $p=0.7$) always lies above the solid brown curve (for three-level engine), which implies that noise-induced coherence enhances the power relative  fluctuations in the engine for entire range of $\lambda$. Thus we can conclude that depending on the parametric regime of operation, the phenomenon of noise-induced coherence can either suppress or enhance the relative power fluctuations in degenerate maser heat engine as compared to its nondegenerate counterpart.

Just like the three-level  engine, the TUR quantifier $\mathcal{Q}_{\rm NIC}$ for the degenerate four-level engine remains invariant under the transformations  $\Gamma_c\rightarrow k \Gamma_c$, $\Gamma_h\rightarrow k \Gamma_h$ and $\lambda\rightarrow k \lambda$,
\be
\mathcal{Q}_{\rm NIC}(k \Gamma_h, k \Gamma_c, k \lambda, n_h, n_c, p)  =\mathcal{Q}_{\rm NIC}(\Gamma_h, \Gamma_c, \lambda, n_h, n_c, p).
\ee

Having discussed the behavior of $\mathcal{Q}_{\rm NIC}$ as a function of matter-field coupling parameter $\lambda$, we move to discuss the behavior of $\mathcal{Q}_{\rm NIC}$ as a function of noise-induced coherence parameter $p$ with all other parameters kept fixed at constant values. It is evident from Fig. \ref{FigNICp} that $\mathcal{Q}_{\rm NIC}$ exhibits a minimum at certain numerical value of $p$ lying between [-1,1].  We note that for $p=-1$, $\mathcal{Q}_{\rm NIC}$ is always greater than 2 regardless of the choice of all system-bath parameters. This can be seen analytically. For $p=-1$, we derive following form of TUR quantifier  $\mathcal{Q}_{\rm NIC}$:
\begin{figure}  
 \begin{center}
\includegraphics[width=8.6cm]{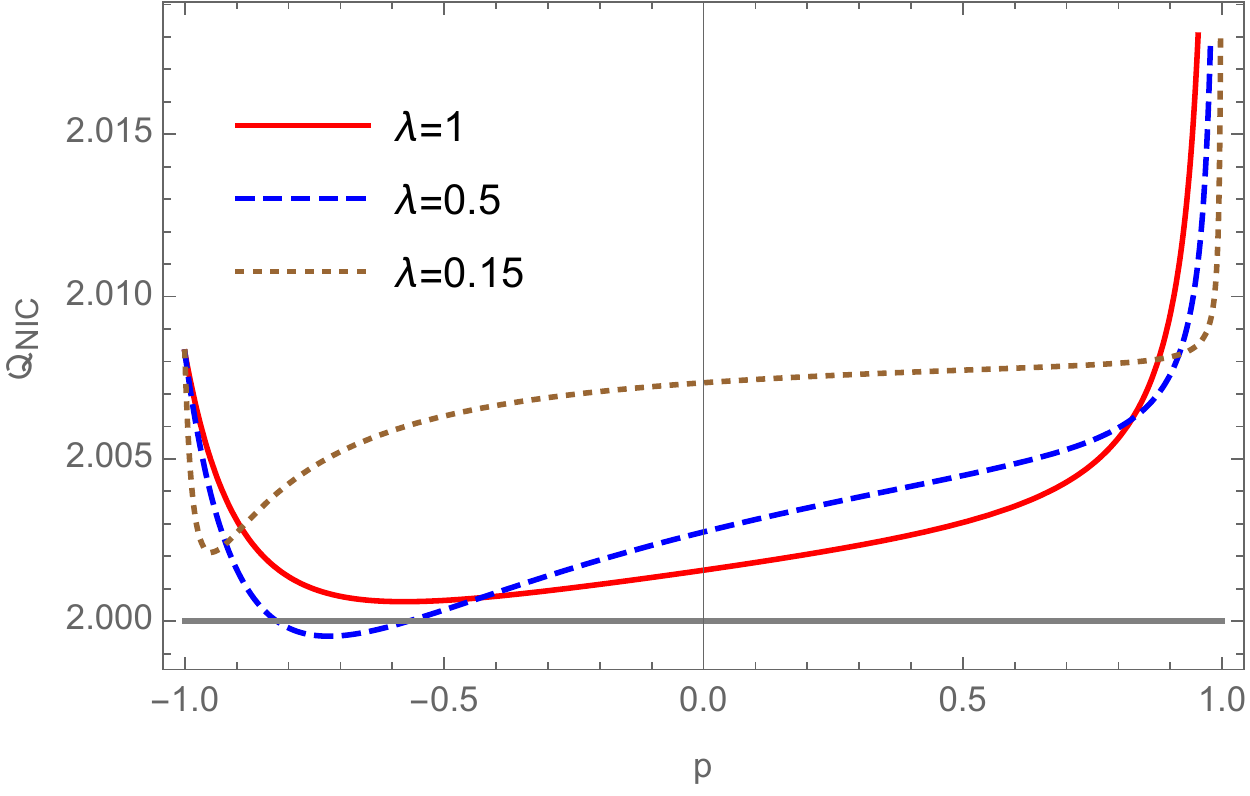}
 \end{center}
\caption{$\mathcal{Q}_{\rm NIC}$    as a function of noise-induced coherence parameter $p$ for different values of $\lambda$. Solid red, dashed blue and   dotted brown curves represent the cases for $\lambda=1$, $\lambda=0.5$ and $\lambda=0.15$, respectively.   Here, $\Gamma_h=0.6$, $\Gamma_c=0.4$, $n_h=5$ and $n_c=2$.
}
\label{FigNICp}
\end{figure}
\be
\mathcal{Q}_{\rm NIC}(p=-1) = \ln\left[ \frac{n_h(n_c+1)}{n_c(n_h+1)}  \right] \frac{n_h+n_c+2n_h n_c}{n_h-n_c},
\ee
which is nothing but the already introduced term $\mathcal{Q}_{\rm pop}$. We have already shown in Sec. III that $\mathcal{Q}_{\rm pop}\ge 2$.
In the ultra-high temperature limit, we have an interesting result. In this case, we can show that $\mathcal{Q}_{\rm NIC}$ is always equal to 2 regardless of the choice of other system-bath parameters:
\be
\mathcal{Q}^{\rm HT}_{\rm NIC} =2.
\ee
This is in contrast to the case with three-level engine. In that case, TUR ratio $\mathcal{Q}_{\rm HT}$ is always less than 2 unless $n_h=n_c$. It implies that in the ultra-high temperature regime, the phenomenon of noise-induced coherence always enhances the relative power fluctuations.

Finally,   we   plot the histogram of sampled values of $\mathcal{Q}_{\rm NIC}$ for randomly sampling over a region of the parametric space (see Fig. 3 (c). Similar to the three level case, for the great majority of sampled operational points, the TUR ratio stays close to 2. However, the minimum value of $\mathcal{Q}_{\rm NIC}$ extracted from histogram (see Inset of Fig. 3 (c)) is 1.985 which is greater than the corresponding case of three-level engine (Model I), i. e., $\mathcal{Q}^{\rm I}_{\rm min}<\mathcal{Q}^{\rm II}_{\rm NIC}$. 

\section{Conclusions}
\label{sec:conclusions}
We have presented a detailed analysis of thermodynamic uncertainty relations in nondegenerate three-level and degenerate four-level maser heat engines. 
For the nondegenerate three-level maser heat engine, we studied two slightly different configurations of the engine and obtained analyical expressions  for the TUR ratio.   We showed that the invariance of TUR quantifier $\mathcal{Q}$   with respect to  scaling of the matter-field  and system-bath coupling constants. Further, for the degenerate four level engine, we studied the effects of noise-induced coherence on TUR. We showed that depending on the parametric regime of operation, the phenomenon of noise-induced coherence can either suppress or enhance the relative power fluctuations.

As noticed in Section \ref{sec:TUR_3_level}, the violation of STUR might be traced back to contributions coming from coherence.
Hence, it would be very interesting to calculate the exact contribution of coherences in the violation of standard TUR. 
To this end, it is necessary to go beyond steady state and use the quantum trajectory approach to unravel the master equation \cite{VuSaito2022}.
We hope to address this point in the near future.

\section{Acknowledgements} 
Varinder Singh, Vahid Shaghaghi and Dario Rosa acknowledge support by the Institute for Basic Science in Korea (IBS-R024-D1). We thank Jae Sung Lee for interesting discussions.

 \onecolumngrid

\appendix\section{Density matrix equations for three-level maser heat engine}
Here, we will present the density matrix equations for the two different variants of the SSD engine. 
\subsection*{Model I}
For three-level system shown in Fig. 1(a), the time evolution of the elements of the density matrix are governed by following equations \cite{Dorfman2018,Varinder2020,BoukobzaTannor2007}:
\begin{eqnarray}
\dot{\rho}_{gg} &=&   \Gamma_h(n_h+1)\rho_{11}+\Gamma_c(n_c+1)\rho_{00}   - (\Gamma_h n_h + \Gamma_c n_c) \rho_{gg}, \nonumber
\\
\,
\\
\dot{\rho}_{11} &=& i\lambda (\rho_{10}-\rho_{01}) - \Gamma_h[(n_h+1)\rho_{11}-n_h\rho_{gg}],\label{A1} \\
\dot{\rho}_{00} &=& -i\lambda (\rho_{10}-\rho_{01}) - \Gamma_c[(n_c+1)\rho_{00}-n_c\rho_{gg}], \\
\dot{\rho}_{10} &=&  i\lambda(\rho_{11}-\rho_{00}) -\frac{1}{2}[\Gamma_h(n_h+1)+\Gamma_c(n_c+1)]\rho_{10}  , \nonumber
\\
\\
%
%
\dot{\rho}_{01} &=& \dot{\rho}_{10}^*. \label{A5}
\end{eqnarray}
\subsection*{Model II}
For three-level system shown in Fig. 1(b), the dynamical equations for different density matrix elements are given by:
 \begin{eqnarray}
\dot\rho_{11} &= &\Gamma_h n_h\rho_{gg} +\Gamma_c n_c\rho_{00}-[\Gamma_h(n_h+1)+\Gamma_c(n_c+1)] \rho_{11}, \nonumber
\\
\vspace{1mm} 
\\
\dot\rho_{00} &=& \Gamma_c (n_c+1)\rho_{11} - \Gamma_c n_c \rho_{00} + \imai \lambda (\rho_{0g}-\rho_{g0})\label{Eqrho00} ,
\\
\dot\rho_{gg} &=& \Gamma_h (n_h+1)\rho_{11} - \Gamma_h n_h\rho_{gg} -\imai\lambda (\rho_{0g}-\rho_{g0})\label{Eqrholl}\,,
\\
\dot\rho_{g0} &=&  \imai\lambda(\rho_{00}-\rho_{gg})-\frac{1}{2}(\Gamma_h n_h+\Gamma_c n_c) \rho_{g0}\,, 
\\
\dot\rho_{0g} &=& \dot\rho^*_{g0}  \label{Eqrhoul}
\end{eqnarray}

\section{Master Equation for four-level maser heat engine}
Consider a modified version of the SSD engine where we have replaced a single upper level $\vert 1\rangle$ by a pair of two  degenerate states $\vert 1\rangle$ and  $\vert 2\rangle$. Then bare Hamiltonian of the four-level system and semiclassical system-field interaction Hamiltonian  are given by \cite{Dorfman2018}
\begin{align}\label{eq:H01}
	H_0=\hbar\sum_{i=g,0,1,2}\omega_i|i\rangle\langle i|,
\end{align}

\begin{align}
	V(t)=\hbar\lambda[e^{-i\omega t}(|1\rangle\langle 0|+|2\rangle\langle 0|)+e^{i\omega t}(|0\rangle\langle 1|+|0\rangle\langle 2|)].
\end{align}
In a rotating frame with respect to $H_0$, the time  evolution of the system is described by the following  master equation:
\begin{equation}
\dot{\rho} = -\frac{i}{\hbar} [H_0+V(t),\rho] + \mathcal{L}_{h}[\rho] + \mathcal{L}_{c}[\rho],
\end{equation}
where $\mathcal{L}_{h(c)}$ represents the dissipative Lindblad superoperator describing the system-bath  interaction with the hot (cold) reservoir:
\begin{eqnarray}
\mathcal{L}_c[\rho]= \Gamma_c(n_c+1) \Big(A_c\rho A_c^\dagger  -\frac{1}{2}\big\{ A_c^\dagger A_k,\rho \big\}\Big) 
+  \Gamma_c n_c\Big(A_c^\dagger \rho A_c  -\frac{1}{2}\big\{ A_c A_c^\dagger,\rho \big\}\Big)  , \label{disscold}
\end{eqnarray}
\begin{eqnarray}
\mathcal{L}_h[\rho]=\sum_{k={1,2}}\Gamma_{hk}\Big[(n_h+1)\Big(A_k\rho A_k^\dagger  -\frac{1}{2}\big\{ A_k^\dagger A_k,\rho \big\}\Big) 
+  n_h\Big(A_k^\dagger \rho A_k  -\frac{1}{2}\big\{ A_k A_k^\dagger,\rho \big\}\Big)\Big] \nonumber
\\
+  \Gamma\cos\theta \Big[(n_h+1)\Big(A_1\rho A_2^\dagger -\frac{1}{2}\big\{ A_2^\dagger A_1,\rho \big\}\Big) 
 +
n_h\Big(A_1^\dagger \rho A_2  -\frac{1}{2}\big\{ A_2 A_1^\dagger,\rho \big\}\Big)\Big] \nonumber
\\
+  \Gamma\cos\theta \Big[(n_h+1)\Big(A_2\rho A_1^\dagger -\frac{1}{2}\big\{ A_1^\dagger A_2,\rho \big\}\Big) 
 +
n_h\Big(A_2^\dagger \rho A_1  -\frac{1}{2}\big\{ A_1 A_2^\dagger,\rho \big\}\Big)\Big]. \label{disshot}
\end{eqnarray} 
where  $A_c=\vert g\rangle \langle 0\vert$,  
$A_k=\vert g\rangle \langle k \vert$ ($k=1, 2$) are jump operators between the relevant transitions.
The time evolution of the density matrix equations is given by
\begin{eqnarray}
	\dot{\rho}_{11}&=& i\lambda(\rho_{10}-\rho_{01})-\Gamma_{h}[(n_{h}+1)\rho_{11}-n_{h}\rho_{gg}] -\frac{1}{2} p \Gamma_h (n_h+1)(\rho_{12}+\rho_{21}),
\\
	\dot{\rho}_{22}&=& i\lambda(\rho_{20}-\rho_{02})-\Gamma_{h}[(n_h+1)\rho_{22}-n_h\rho_{gg}] \frac{1}{2} p  \Gamma_h (n_h+1)(\rho_{12}+\rho_{21}),
\\
	\dot{\rho}_{00}&=& i\lambda(\rho_{01}+\rho_{02}-\rho_{10}-\rho_{20})\notag - \Gamma_c[(n_c+1)\rho_{00}-n_c\rho_{gg}],
\\
	\rho_{gg} &=& 1-\rho_{11}-\rho_{22}-\rho_{00},
\\
	\dot{\rho}_{12}&=&i\lambda(\rho_{10}-\rho_{02})-\frac{1}{2}[\Gamma_{h}(n_h+1)+\Gamma_{h}(n_h+1)]\rho_{12}
-\frac{1}{2} p \Gamma_h[(n_h+1)\rho_{11}
	 +(n_h+1)\rho_{22}-(n_h+n_h)\rho_{gg}], \nonumber
	 \\
\\
	\dot{\rho}_{10}&=&i\lambda(\rho_{11}-\rho_{00}+\rho_{12})-\frac{1}{2}[\Gamma_c(n_c+1)+\Gamma_h(n_h+1)]\rho_{10}	-\frac{1}{2} p \Gamma_h (n_h+1)\rho_{20},
\\
	\dot{\rho}_{20}&=&i\lambda(\rho_{22}-\rho_{00}+\rho_{21})-\frac{1}{2}[\Gamma_c(n_c+1)+\Gamma_{h}(n_h+1)]\rho_{20}
	-\frac{1}{2} p \Gamma_h(n_h+1)\rho_{10}.
\end{eqnarray}

\section{Full Counting Statistics}
\label{AppFCS}

In order to calculate $\mathcal{Q}$, we first need to calculate mean and variance of the power along with the rate of entropy production. For the steady state heat engines obeying strong-coupling condition (no heat leaks between the reservoirs), the relation between the energy flux (heat and work fluxes) $I_E$ and matter flux (here photon flux) $I$ is given by \cite{Esposito2009}:
\be
I_E = \epsilon I.
\ee
 The above equation implies that the energy is transported  by the particles of a given energy $\epsilon$. In case of the work flux (power), the above relation reduces to $P=(\omega_h-\omega_c)I$. Similarly, the variance of the power is given by $\text{var} (P)=(\omega_h-\omega_c)^2 \text{var} (I)$. Using these relations, the TUR ratio $\mathcal{Q}$ can be written as
 \be
 \mathcal{Q} =   \sigma \frac{ {\rm var} (P)}{P^2}= \sigma \frac{ {\rm var} (I)}{I^2}. \label{Aux1}
 \ee
Further, for the three-level heat engine, $\sigma$, entropy production rate assumes the following form \cite{BoukobzaTannor2007}
\be
\sigma = \ln \left[ \frac{n_h(n_c+1)}{n_c(n_h+1)}\right] I > 0. \label {EP}
\ee
Using Eq. (\ref{EP}) in Eq. (\ref{Aux1}), we have
\be
\mathcal{Q} =   \ln \left[ \frac{n_h(n_c+1)}{n_c(n_h+1)}\right] \frac{{\rm var} (I)}{I}.  \label{TURaux}
\ee

Now we will evaluate the expression for the photon flux $I$. In open quantum systems, the particle statistics can be determines by using the formalism of full counting statistics (FCS), where counting fields are incorporated in the master equation. For our purpose, it is sufficient to introduce counting field either for the hot or the cold reservoir. Here, we choose to introduce the counting field ($\chi$) for the cold reservoir.  \cite{Bruderer2014,SchallerBook}. The modified Lindblad master equation takes the following form
\begin{equation}
\dot{\rho} = -\frac{i}{\hbar} [V_R,\rho] + \mathcal{L}_{h}[\rho] + \mathcal{L}^{\chi}_{c}[\rho],
\end{equation}
where  the modified Lindblad superoperator
%
\be
\mathcal{L}_c[\rho] = \Gamma_c(n_c+1) \big (\e^{-i\chi} \sigma_{g0} \rho \sigma^{\dagger}_{g0} -\frac{1}{2}  \{\sigma^{\dagger}_{g0} \sigma_{g0},\rho\}   \big) +
\Gamma_c n_c  \big (\e^{i\chi}  \sigma^{\dagger}_{g0} \rho \sigma_{g0} -\frac{1}{2}  \{\sigma_{g0} \sigma^{\dagger}_{g0} ,\rho\}   \big). \label{D2A}
\ee
By  vectorizing the  density matrix elements into a state vector $\rho_R=(\rho_{gg}, \rho_{00}, \rho_{11}, \rho_{10}, \rho_{01})^T$, we can write  the  above Lindblad master equation as a matrix equation with the  Liouvillian supermatrix $\mathcal{\textbf{L}}(\chi)$
\begin{equation}
\dot{\rho} = \mathcal{\textbf{L}}(\chi) \rho,
\end{equation}
where
\begin{equation}
\mathcal{L}(\chi) = \left[
\begin{matrix}
-(\Gamma_h n_h + \Gamma_c n_c) & \Gamma_c (n_c+1)\e^{-i \chi} & \Gamma_h (n_h+1) & 0 & 0
\\
\Gamma_c n_c \e^{i\chi} & -\Gamma_c(n_c+1) & 0 & -i\lambda & i\lambda
\\
\Gamma_h n_h  & 0& -\Gamma_h(n_h+1) & i\lambda  & -i\lambda
\\
0 & -i\lambda & i\lambda & -\frac{1}{2}[\Gamma_h(n_h+1)+\Gamma_c(n_c+1)] & 0I
\\
0 & i\lambda & -i\lambda & 0 &  -\frac{1}{2}[\Gamma_h(n_h+1)+\Gamma_c(n_c+1)]  \\
\end{matrix}
\right]\,.
\label{EqFCSLiouvillian}
\end{equation}
For $\chi\rightarrow0$, Eq. (\ref{EqFCSLiouvillian}) reduces to the original Liouvillian operator (given in Eq. (4)) for standard time evolution. 

In the long time limit, the $k$'th cumulant of the integrated number of quanta (number of photons here) emitted into the cold reservoir  can be determined by \cite{SchallerBook}
\begin{equation}
\label{eq:cumulants}
C^k(t) =\left.(\imai\partial_{\chi})^k\left[\xi(\chi) \right]\right|_{\chi=0}\,,
\end{equation}
where $\xi(\chi)$ is the eigenvalue of $\mathcal{L}(\chi)$ with the largest real part. The first cumulant corresponds to the mean current and the second cumulant corresponds to the variance:
\begin{equation}
I \simeq \left. \imai \partial_{\chi}\xi(\chi)\right|_{\chi = 0}\,,\hspace{1.5cm}
\mathrm{var}(I) \simeq\left. -\partial_{\chi}^2\xi(\chi)\right|_{\chi = 0}.
\label{EqTURCumulant}
\end{equation}

To obtain the expressions for the mean and variance, we follow the method explained in Ref.~\cite{Bruderer2014}.
Consider the characteristic polynomial of $\mathcal{L}(\chi)$
\begin{equation}
\sum_n c_n \xi^n = 0\,, \label{sum}
\end{equation}
where the terms $c_n$ are functions of $\chi$.
Define
\begin{equation}
c_n' = \imai\partial_{\chi} c_n|_{\chi=0},
\quad c_n'' = (\imai\partial_{\chi})^2c_n|_{\chi=0} = -\partial_{\chi_u}^2c_n|_{\chi=0}\,.
\end{equation}
Differentiating Eq. (\ref{sum}) with respect to the counting parameter $\chi$, and then evaluating the resulting equation at $\chi=0$, we have
\begin{equation}
\left[\imai\partial_{\chi}\sum_n c_n\xi^n\right]_{\chi=0} = \sum_n[c_n'+(n+1)c_{n+1}\xi']\xi^n(0) = 0\,.
\label{EqFCSfirst}
\end{equation}
By taking the  second order derivative of Eq. (\ref{sum}), we find
\begin{equation}
\left[(\imai\partial_\chi)^2\sum_nc_n\xi^n\right]_{\chi=0} =
\sum_n[c_n''+2(n+1)c_{n+1}'\xi'+(n+1)c_{n+1}\xi''+(n+1)(n+2)c_{n+2}\xi'^2]\xi^n(0)=0\,.
\label{EqFCSsecond}
\end{equation}
As the zeroth term $\xi^0=1$ should vanish, hence Eq.~\eqref{EqFCSfirst} implies
\begin{equation}
c_0'+c_1\xi'=0\,,
\end{equation}
from which we obtain the expression for the current
\begin{equation}
I = \xi' = -\frac{c_0'}{c_1}\,.
\label{EqFCSmean}
\end{equation}
Similarly from Eq.~\eqref{EqFCSsecond}, we obtain the following expression for the variance
\begin{equation}
\mathrm{var}(I) = \xi'' =-\frac{c_0''+2I(c_1'+c_2I)}{c_1} =  2\frac{c_0'c_1c_1'-c_0'^2c_1'}{c_1^3}-\frac{c_0''}{c_1}.
\label{EqFCSvariance}
\end{equation}

Applying the above-mentioned procedure to the Liouvillian given in Eq.~\eqref{EqFCSLiouvillian}, we obtain 
\begin{equation}
\begin{split}
c_0'=& (n_h-n_c) \Gamma_h\Gamma_c  \Gamma' \lambda^2,
\\
c_0'' =&   (2n_h n_c+n_h+n_c)\Gamma_h\Gamma_c  \Gamma'  \lambda^2,
\\
c_1 =&  \frac{1}{4}\Gamma'      \big[  (3n_h n_c +2n_h+2n_c+1)\Gamma_c\Gamma_h  \Gamma'+4[\Gamma_h(3n_h+1)+\Gamma_c(3n_c+1)]                            \big],
\\
c_1' =& 2(n_h-n_c)\Gamma_h\Gamma_c\lambda^2,
\\
c_2 =& -\frac{1}{4} \Big\{  (n_h+1)^2 (2n_h+1)\Gamma_h^3    +  (n_c+1)^2 (2n_c+1)\Gamma_c^3   +  (n_h+1)[7+13n_h+6(2+3n_h)n_c]
\\
+& (n_c+1)[7+13n_c+6(2+3n_c)n_h] \Big\} - 4 [(2n_h+1)\Gamma_h+(2n_c+1)\Gamma_c]
\end{split}
\end{equation}
where $\Gamma'=\Gamma_h (n_h+1)+\Gamma_c (n_c+1)$. This provides the current
\begin{equation}
I =  \frac{4(n_h-n_c)\Gamma_h\Gamma_c\lambda^2}
{
4\lambda^2[\Gamma_h(3n_h+1)+\Gamma_c(3n_c+1)] + (3n_h n_c+2n_h+2n_c+1)[\Gamma_h(n_h+1)+\Gamma_c(n_c+1)]\Gamma_h\Gamma_c.
}
\label{EqFCScurrent}
\end{equation}
In the similar manner, using Eq. (\ref{EqFCSvariance}) we can obtain the expression  for $\text{var}(I)$, and further using this expression in Eq. (\ref{TURaux}), we   finally obtain the expression for TUR ratio $\mathcal{Q}^{\rm I}$ as given by Eq. (\ref{TURme}) in the main text.

By following the same procedure, we can obtain the expressions for  $\mathcal{Q}^{I\rm I}$ (Eq. \ref{TURpatrick})  and $\mathcal{Q}_{\rm NIC}$. The analytic expression for  $\mathcal{Q}_{\rm NIC}$ is rather complicated and not illuminating at all. Hence, we will not present it here.  In the main text, we have provided the expressions of $\mathcal{Q}_{\rm NIC}$ for certain specific cases.
 
 \twocolumngrid

 \bibliography{TUR_NIC}

\bibliographystyle{apsrev4-2}

\end{document}